\documentclass[jkps,twoside,twocolumn,fleqn,showpacs,showkeys,showdoi]{revtex4}
\usepackage{graphicx}
\usepackage{amssymb}
\usepackage{amsmath}
\usepackage{bm}
\newcommand{\bro}{BaRuO$_3$}

\begin{document}
\setcounter{page}{0}
\title[]{Effects of Magnetovolume and Spin-orbit Coupling in the Ferromagnetic 
 Cubic Perovskite BaRuO$_3$}
\author{Young-Joon \surname{Song}}
\affiliation{Department of Applied Physics, Graduate School, Korea University,
 Sejong 339-700, Korea }
\author{Kwan-Woo \surname{Lee}}
\email{mckwan@korea.ac.kr}
\affiliation{ Department of Display and Semiconductor Physics, Korea University,
 Sejong 339-700, Korea \\
Department of Applied Physics, Graduate School, Korea University,
 Sejong 339-700, Korea }
\date{\today}

\begin{abstract}
\bro~ having five different crystal structures has been synthesized by
varying the pressure while sintering.
Contrary to the other phases being nonmagnetic, 
the cubic perovskite phase synthesized recently shows an itinerant ferromagnetic character.
We investigated this ferromagnetic \bro~ using first principles calculations.
A few van Hove singularities appear around the Fermi energy, causing
unusually high magnetovolume effects of $\Delta M/\Delta a \approx$ 4.3 $\mu_B$/\AA~
as well as a Stoner instability [$IN(0)\approx$ 1.2].
At the optimized lattice parameter $a$, the magnetic moment $M$ is 1.01 $\mu_B$ 
in the local spin density approximation. 
When spin-orbit coupling is included, the topologies of some Fermi surfaces 
are altered, and the net moment is reduced by 10\% to a value
very close to the experimentally observed value of $\sim$ 0.8 $\mu_B$.
Our results indicate that this ferromagnetism is induced by the Stoner instability,
but the combined effects of the $p-d$ hybridization, the magnetovolume, and 
the spin-orbit coupling determine the net moment.
In addition, we briefly discuss the results of the tight-binding Wannier function technique.
\end{abstract}



\maketitle

\section{INTRODUCTION}
Ruthenates, which formally possess tetravalent $4d^4$ Ru ions, show
interesting properties: unconventional superconductivity in Sr$_2$RuO$_4$,\cite{sr2ruo4}
a metal-insulator transition in La$_2$Ru$_2$O$_7$ and Li$_2$RuO$_3$,\cite{la2ru2o7,li2ruo3}
metamagnetism in Sr$_3$Ru$_2$O$_7$,\cite{sr3ru2o7} and complicated magnetic behaviors
in the double pyrochlore $RE$$_2$Ru$_2$O$_7$ ($RE$=rare-earth element)\cite{ru2o7}. 
In particular, the Ru-based perovskites {$\cal A$}RuO$_3$ ({$\cal A$}=alkaline-earth elements or Pb)
have been studied extensively for several decades 
due to their atypical and controversial electronic and magnetic 
behaviors\cite{good68,mazin,cava98,rabe,good08prl,jung, cava}.
SrRuO$_3$ is a metallic ferromagnet with T$_C=$ 160 K and an experimentally
observed moment of $\sim$ 1.6 $\mu_B$, whereas CaRuO$_3$ is a nonmagnetic metal\cite{shirako,klein}.
PbRuO$_3$ seems to exhibit incipient magnetism, but whether this system is a metal or an insulator 
at low-T is still under debate\cite{attfield,good09}.
The crystal structure of BaRuO$_3$ depends strongly on
the temperature and pressure during synthesis\cite{good08} 
whereas the other compounds are orthorhombic.
Depending on the amount of corner- and face-sharing RuO$_6$, the structures of \bro~ 
are categorized as the four-, six-, and ten-layered hexagonal phases
($4H$, $6H$, and $10H$, respectively), in addition to the nine-layered rhombohedral
phase ($9R$)\cite{10h,9r,6h4h,4h,6h}.
As discussed both theoretically and experimentally\cite{10h,felser,cava},
the $10H$ and the $9R$ phases seem to be semi-metallic 
whereas the rest are nonmagnetic metals.
Early in 2008, Jin {\it et al.} synthesized the cubic phase 
using a high T ($\sim$ 1000 $^\circ$C) and high P ($\sim$ 18 GPa) technique\cite{good08}.  

In this research, we will focus on the cubic phase, which shows ferromagnetic behavior 
with T$_C=$ 60 K\cite{good08}.
On the basis of resistivity measurements, which show a sharp peak in the T-derivation of
the resistivity at T$_C$, Jin {\it et al.} suggested 
simple metallic characteristics in the high-T region, but a Fermi-liquid behavior 
in the low-T regime.
The effective magnetic moment is 2.6 $\mu_B$, close to the value for $S=1$.
However, the saturated moment of 0.8 $\mu_B$ is significantly less than 
the value corresponding to $S=1$, implying atypical magnetic behaviors.
Moreover, when pressure is applied, $T_C$ decreases, reaching 50 K at $P=3$ GPa.

The elastic properties and the pressure effects have been theoretically investigated 
to some extent\cite{cai,han}, but no detailed studies of the magnetic 
and electric properties of this cubic phase have been reported yet.
Thus, we will address the magnetic properties of this system
through first-principles calculations, including the local spin density approximation (LSDA),
the LSDA plus spin-orbit coupling (LSDA+SOC), the fixed spin moment (FSM), 
and a tight-binding Wannier analysis.

\section{Structure and Method of Calculation}
First, we optimized the lattice parameter $a$ in the cubic phase.
LSDA calculations lead to our optimized $a=$ 3.938 \AA. 
This value is about 1.7\% smaller than the experimentally-observed value
at room temperature\cite{good08}. 
We used the optimized lattice parameter in our calculations, 
since the total moment at the optimized $a$ 
is close to the experimental value, as will be discussed in the next section.

Our calculations were carried out with the LSDA and the LSDA+SOC
implemented in the all-electron full-potential code WIEN2k\cite{wien2k}.
The basis size was determined by using R$_{mt}$K$_{max}=7$ 
and augmented plane wave sphere radii of
2.5 for Ba, 2.0 for Ru, and 1.7 for O.
The Brillouin zone was sampled with a 14$\times$14$\times$14 $k$-mesh.
For both the LSDA+SOC and the FSM calculations\cite{fsm}, 
a much denser $k$-mesh of 20$\times$20$\times$20 was used.
The lattice parameter was relaxed until the forces were smaller than
2 mRy/a.u.
Furthermore, the all-electron full-potential local-orbital code FPLO was used for
the tight-binding Wannier function analysis\cite{fplo}.

\section{Results}
\begin{figure}[tbp]
{\resizebox{8cm}{6cm}{\includegraphics{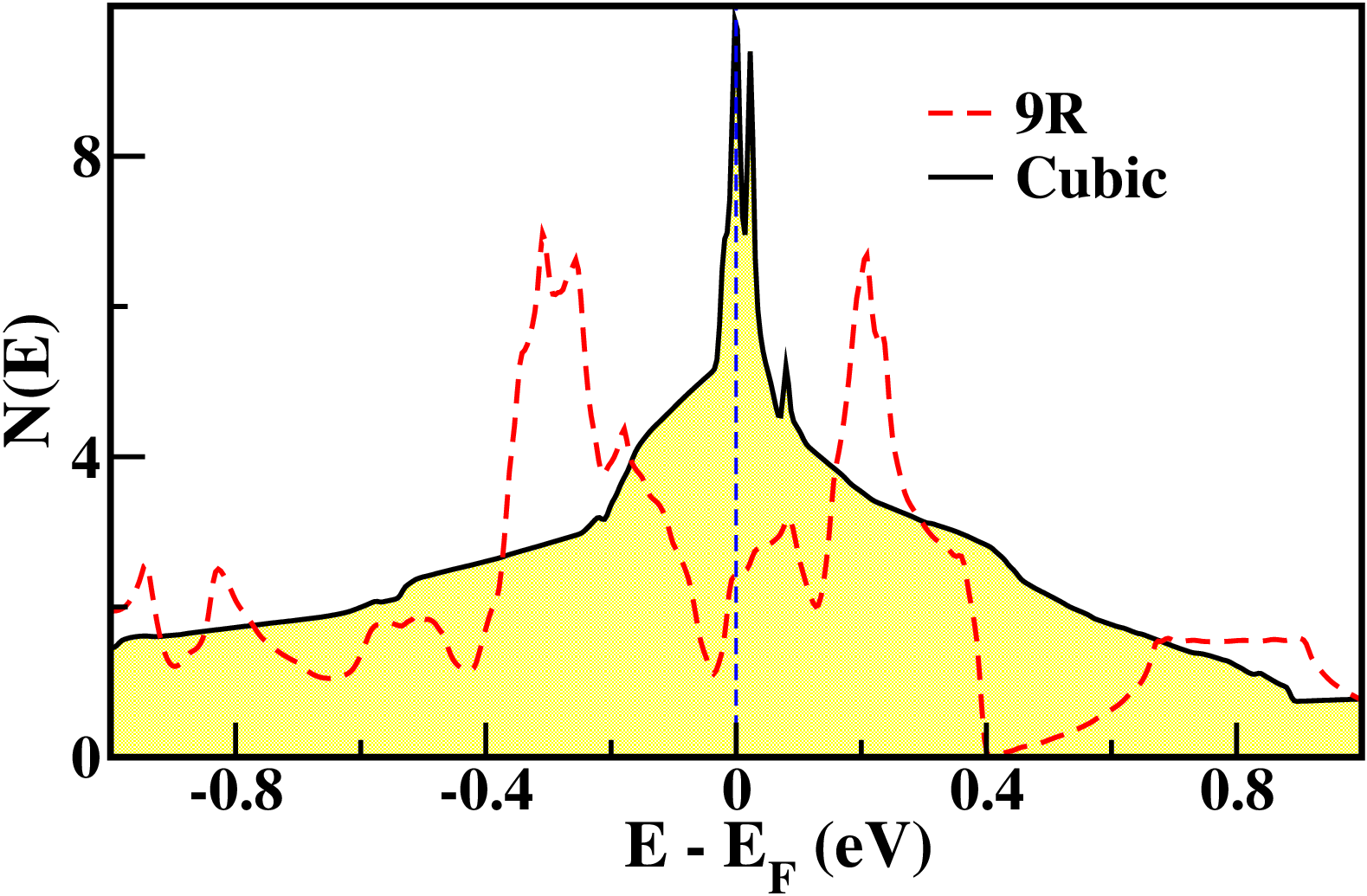}}}
\caption{(Color online) Comparison of the nonmagnetic total densities of states (DOSs)
per formula unit between cubic and $9R$ \bro.
For the cubic phase, the optimized lattice parameter was used.
In the cubic phase, two sharp peaks appear around the Fermi energy $E_F$,
resulting in a strong magnetic instability.
The vertical dashed line indicates $E_F$, which is set to zero.
}
\label{pmdos}
\end{figure}

\begin{figure}[tbp]
\vskip 2mm
{\resizebox{8cm}{6cm}{\includegraphics{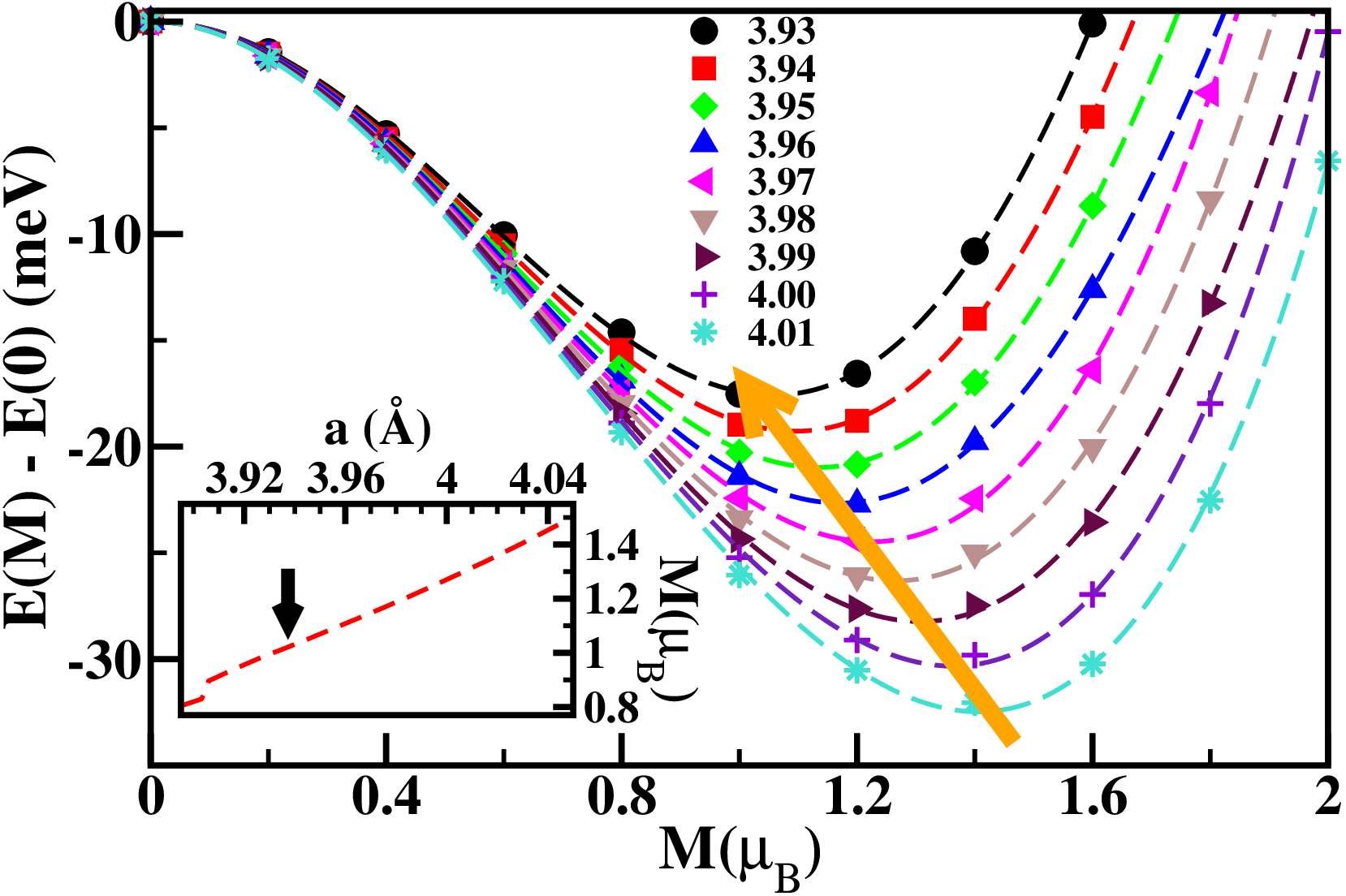}}}
\caption{(Color online) FSM calculations in the cubic phase, 
while changing the lattice parameter $a$ (in units of \AA). 
$E(0)$ denotes the energy of the nonmagnetic state for each $a$. 
The arrow roughly connects the minimum energy states.
{\it Inset}: Change in the total magnetic moment $M$ with respect to $a$ 
in the FM calculations. The arrow indicates the value at the optimized $a$ .
A discontinuity occurs at $a\sim$ 3.91 \AA, because $E_F$ pins 
a van Hove singularity in the spin-up channel (see text).
}
\label{fsm}
\end{figure}

\begin{figure}[tbp]
{\resizebox{8cm}{6cm}{\includegraphics{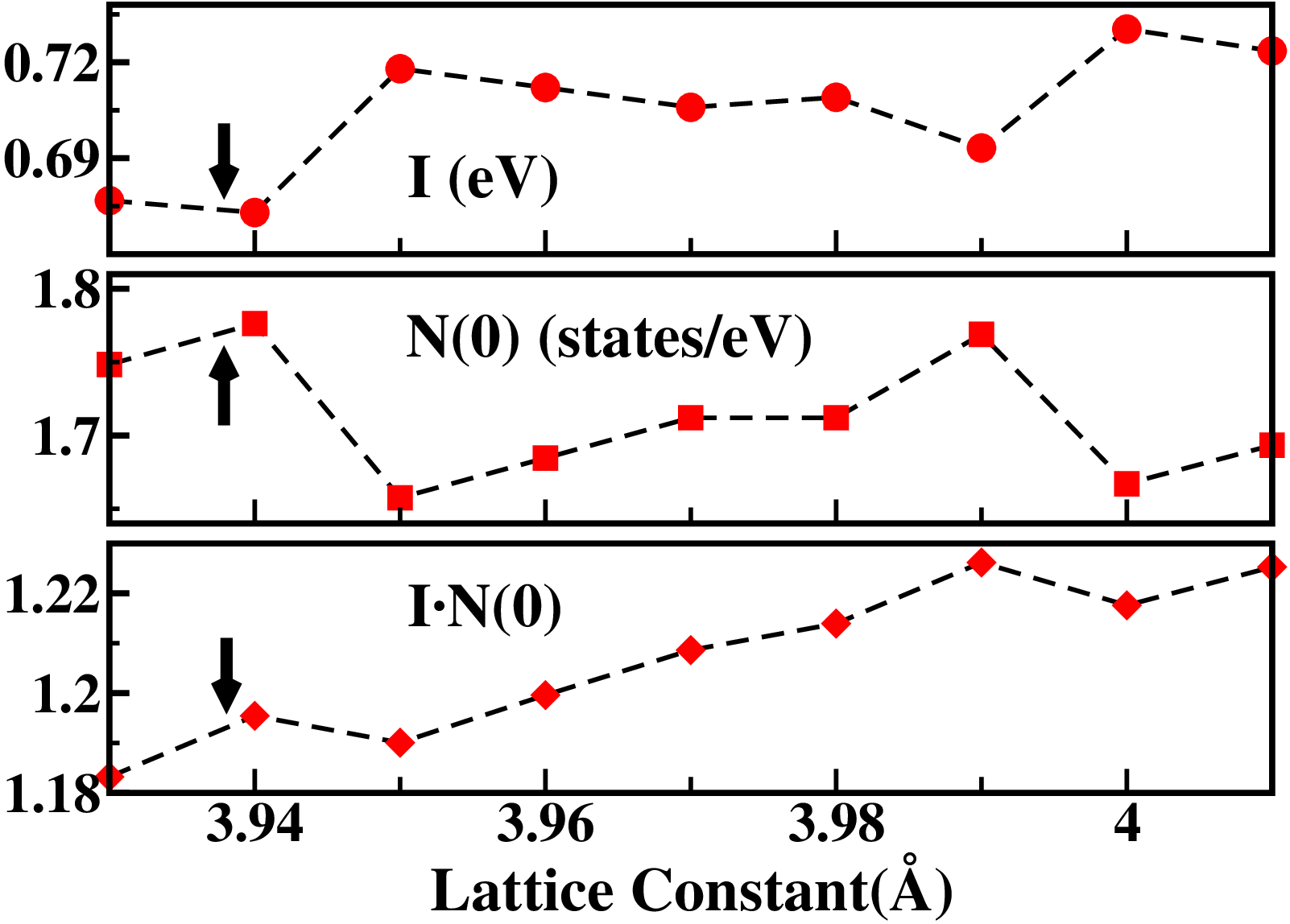}}}
\caption{(Color online) Change in the Stoner parameter $I$ (top),
the DOS $N(0)$ per single spin at $E_F$ (middle), and $IN(0)$ (bottom) 
with the lattice parameter in the cubic \bro.
Roughly, $IN(0)$ decreases linearly as the lattice parameter decreases.
}
\label{iv}
\end{figure}

Figure \ref{pmdos} shows the overlapped total densities of states (DOSs) 
between the nonmagnetic $9R$ and cubic phases in the range of --1 to 1 eV.
The DOS of the cubic phase is five times larger than that of the $9R$ phase 
due to the sharp peaks at $E_F$ and 30 meV, suggesting strong magnetic 
instability.

Using the LSDA, we calculated the change in the total magnetic moment
while varying $a$ from the experimental value to a value slightly smaller
than the optimized one.
As shown in the inset of Fig. \ref{fsm}, the magnetic moment decreases linearly
with a slope of $\Delta M/\Delta a\approx$ 4.3 $\mu_B$/\AA,
indicating strong magnetovolume effects.
This value is much larger than that of $\sim$ 1 $\mu_B$/\AA~ in the Invar systems 
${\cal A}$Fe$_2$ ({$\cal A$}=Y, Zr, Hf, and Lu)\cite{tu}. 

To clarify these behaviors, FSM calculations were conducted.
Plots of energy versus total magnetic moment $M$ for varying values of 
$a$ are shown in Fig. \ref{fsm}.
The difference in energy between the $M=$0 and the ferromagnetic (FM) ground states
decreases monotonically from $\sim$ 32 meV at $a=$4.01 \AA~ to 17 meV at 3.93 \AA.
The curves are fitted well by $E(M)-E(0) \approx -\alpha M^2 + \beta M^4$,
with constants $\alpha$ and $\beta$, in the low-$M$ region. 
For the bare susceptibility $\chi_0=2\mu_B^2N(0)$, the Stoner-enhanced 
susceptibility is given by $\chi=\chi_0/[1-N(0)I]$, where $N(0)$ is the single-spin
DOS at $E_F$. In the low-$M$ region, $\alpha=\frac{1}{2}\chi^{-1}$.
The top panel of Fig. \ref{iv} displays the obtained Stoner parameters
$I \approx$ 0.70($\pm 0.02$) eV for each $a$. 
As $a$ changes, the DOS at $E_F$ fluctuates substantially owing to 
van Hove singularities (vHSs) near $E_F$ (see below), 
as given in the middle panel of Fig. \ref{iv}.
$IN(0)$ decreases monotonically as $a$ decreases, 
which is consistent with the substantial magnetovolume effects,
but for all the lattice parameters studied here, 
these values exceed the Stoner instability criterion $IN(0) \ge 1$.

\begin{figure}[tbp]
\vskip 2mm
{\resizebox{8cm}{6cm}{\includegraphics{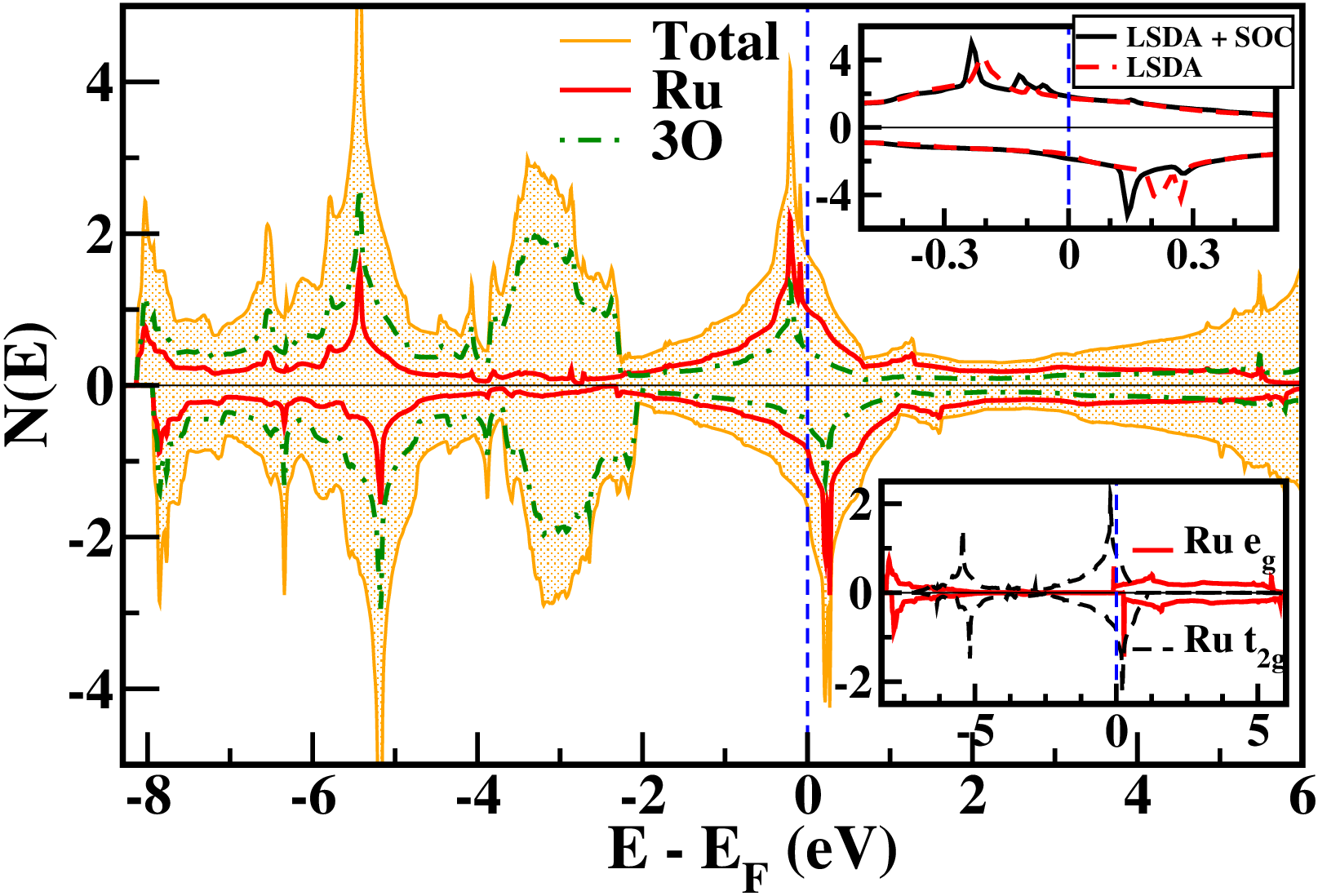}}}
\caption{(Color online) Total and atom-projected DOSs of
the FM cubic \bro~ in the full range, including all O $2p$ and Ru $4d$ states.
In the spin-up channel, peaks appear at --0.2 and --0.1 eV. 
The DOS N(0) at $E_F$ is 3.34 states per eV for both spins.
{\it Upper inset}: Expanded view of the total DOSs in LSDA and LSDA+SOC 
between --0.5 and 0.5 eV.
{\it Lower inset}: Ru $t_{2g}$ and $e_{g}$ orbital-projected DOSs.
}
\label{fmdos}
\end{figure}

\begin{figure}[tbp]
{\resizebox{8cm}{6cm}{\includegraphics{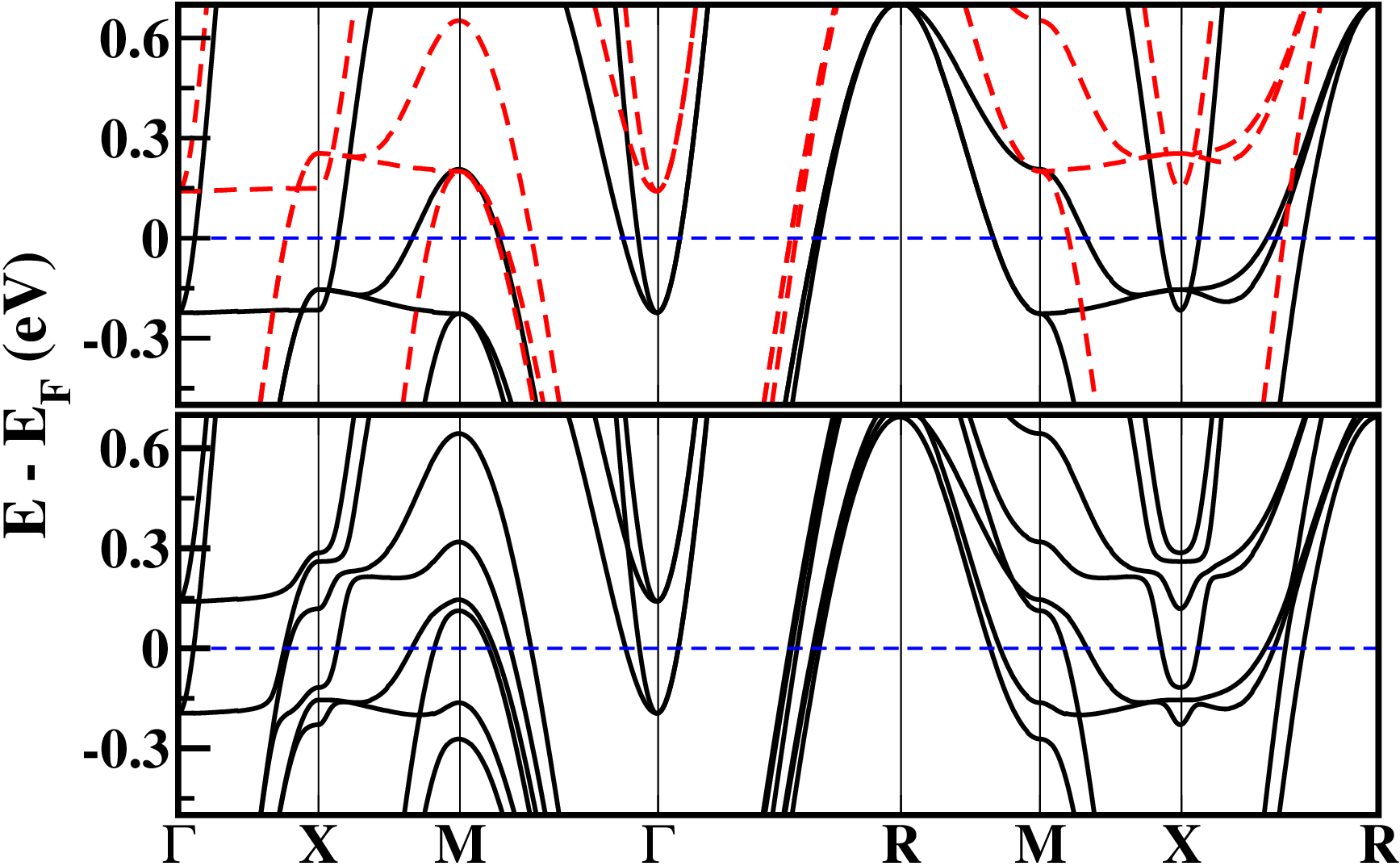}}}
\caption{(Color online) Blowup of the FM band structure in LSDA (top) and LSDA+SOC (bottom)
 near $E_F$ in the cubic \bro.
 In the top panel, the solid and the dashed lines represent the spin-up and -down 
 characters, respectively.
Around the $X$ and the $M$ points, the effects of spin-orbit coupling are substantial.
The horizontal dashed lines indicate $E_F$.
}
\label{fmband}
\end{figure}

Like the FSM results, the LSDA results indicate that 
the FM state is energetically favored over the nonmagnetic state by 16.4 meV 
for the optimized lattice parameter.
For comparison, from a simple Stoner instability, the energy gain
of $Im^2$/4 for $I \approx$ 0.7 eV is 175 meV, much larger than the calculated value.

Now, we will address the electronic structure of the FM states.
Figure \ref{fmdos} shows the total and the atom-projected DOSs, which
can be figured out well using the concept of RuO$_6$ cluster orbitals\cite{mazin}.
The Ru $t_{2g}$ states, hybridized with the O $p_{\pi}$ states,
produce an antibonding manifold at --2 to 1 eV, a bonding manifold
at --6 to --4 eV, and a nonbonding manifold at --4 to --2 eV, 
as can be seen in the lower inset of Fig. \ref{fmdos}.
The bonding and the antibonding states of the Ru $e_g$ and 
the O $p_{\sigma}$ states
lie in the regions of --8 to --6 eV and 0 to 6 eV, respectively.
The exchange splitting of the $t_{2g}$ manifold is about 0.5 eV.
In the FM state, the total moment of 1.01 $\mu_B$ is decomposed into
0.6 $\mu_B$ for Ru and 0.28 $\mu_B$ for 3O. (The remnant is in the interstitial region.)

The enlarged FM band structure in the LSDA is given in the top panel of Fig. \ref{fmband}.
In this region, the spin-up and -down structures are nearly identical, 
except for a difference in the on-site energy of roughly 0.5 eV due to exchange splitting.
In the spin-up channel, in addition to a flat band at --0.2 eV along the $\Gamma$-$X$ line,
which commonly appears in conventional perovskites, 
two saddle points, one each at the $X$ and the $M$ points, occur, resulting in vHSs
just below $E_F$. These features may cause the unusual electrical properties observed 
experimentally\cite{good08}.

The band structure of the $t_{2g}$ and the $p_{\pi}$ clusters in the range of --8 eV to 2 eV 
was reproduced well using the tight-binding Wannier function technique.
We obtained two important parameters: $pd\pi$ hopping $t_{\pi}=1.23$ eV
and direct oxygen-oxygen hopping $t'_{\pi}=0.17$ eV.
Compared with the tight-binding parameters of the nonmagnetic CaRuO$_3$ 
obtained by Mazin and Singh\cite{mazin}, $t_{\pi}$ is about 10\% smaller, 
but $t'_{\pi}$ is about half as large. 
This is consistent with the 20\% smaller bandwidth of the antibonding $t_{2g}$ manifold 
in \bro, implying a stronger magnetic instability.

SOC affects the Fermi surface (FS) topology in 
isovalent Sr$_3$Ru$_2$O$_7$\cite{tamai}, though its effects are often negligible 
in $4d$ systems. To consider the effects of SOC, we conducted LSDA+SOC calculations.
As shown in the upper inset of Fig. \ref{fmdos}, the changes in the DOS are 
relatively small, leading to a tiny orbital moment of --0.013 $\mu_B$ on the Ru ion.
However, the mixing of states with other spin channel reduces the net moment by 10\%,
bringing it closer to the experimentally observed value of 0.8 $\mu_B$\cite{good08}.
The net moment is 0.91 $\mu_B$, with local moments of 0.55 $\mu_B$ for Ru
and 0.26 $\mu_B$ for 3O.

An additional remarkable feature is that vHSs move toward $E_F$
in both spin channels. In the spin-up channel, the vHS at --0.1 eV splits into two peaks
whereas two vHSs merge and move to 0.1 eV in the spin down channel.
These features reflect changes in the band structure at the $X$ and the $M$ points, 
as demonstrated in the enlarged band structure in the bottom panel of Fig. \ref{fmband}.
The strength of the SOC is about 0.1 eV at these points near $E_F$,
but negligible elsewhere.
The SOC alters the FS topology at the $X$-points, in particular, 
in the intersecting pipe-like
and cube-like FSs (not shown here), which are often observed in conventional
perovskites\cite{LP09}. 
The open pipe-like surface becomes closed whereas the cube-like closed one becomes 
open.

\section{DISCUSSION AND SUMMARY}
The Fermi-liquid behavior experimentally observed in the low-T region may 
suggest the presence of correlation effects.
Considering that this material is a metallic $4d$ system, 
the correlation effects should be weak.
In our preliminary calculations using the LDA+U method, 
applying $U$ to the Ru ion shifts the $e_g$ manifold up
and the $t_{2g}$ manifold down,
so only the intersecting pipe-like FS disappears.
To evaluate the importance of the correlation effects,
detailed experimental observations are necessary.

In summary, we investigated the magnetic properties of the ferromagnetic cubic 
perovskite \bro, using various first principles calculations: FSM, LSDA, LSDA+SOC, 
and the Wannier function technique.
In the nonmagnetic DOS, the cubic phase has a sharp peak pinning at $E_F$, 
yielding $IN(0)\approx 1.2$, beyond the Stoner instability.
In the LSDA calculations, a few vHSs appear near $E_F$ in both spin channels,
resulting in a strong magnetovolume effect of $\Delta M/\Delta a\approx$ 4.3 $\mu_B$/\AA,
significantly higher than that of the Invar system ${\cal A}$Fe$_2$.
The effects of the SOC are substantial near $E_F$ around the $X$ and the $M$ points
and affect the topology of some of FSs, 
though the strength of the SOC is negligible in most regions.
At the optimized volume, the effects of the SOC reduce the net moment
by 10\% from the LSDA value, and it becomes very close to 
the experimentally-observed moment.
Our results show that the Stoner instability produces ferromagnetism,
and that the combined effects of the hybridization, the magnetovolume, and the SOC
determine the magnetic moment.
Furthermore, the unusually strong magnetovolume effects may imply atypical 
transport properties  in this FM \bro, suggesting further questions
for experimental research.

\begin{acknowledgments}
This research was supported by Korea University.
\end{acknowledgments}

\end{document}